\pdfoutput=1
\documentclass[fleqn,10pt]{wlscirep}
\usepackage[utf8]{inputenc}
\usepackage[T1]{fontenc}
\usepackage{lineno,hyperref}
\usepackage{amsmath,graphicx,makecell}
\usepackage{amssymb}
\usepackage{amsfonts}
\usepackage{bbm}
\usepackage{changes}
\usepackage{enumitem}
\graphicspath{{./figs/}}
\usepackage{threeparttable}
\usepackage{booktabs}
\usepackage{algorithm}
\usepackage{algpseudocode}
\usepackage{footnote}
\usepackage{caption}
\usepackage{subcaption}
\newcolumntype{C}[1]{>{\centering\arraybackslash}m{1in}}
\newcolumntype{M}[1]{>{\centering\arraybackslash}m{1.88in}}
\newcolumntype{P}[1]{>{\centering\arraybackslash}m{1.2in}}

\title{High Resolution Modeling and Analysis of Cryptocurrency Mining's Impact on Power Grids: Carbon Footprint, Reliability, and Electricity Price}

\author[1]{Ali Menati}
\affil[1]{Department of Electrical and Computer Engineering, Texas A\&M University, College Station, TX, 77843, USA}
\affil[2]{Electric Reliability Council of Texas, Inc., Taylor, TX, 76574, USA}
\affil[3]{Texas A\&M Energy Institute, College Station, TX, 77843, USA}
\author[1]{Xiangtian Zheng}

\author[1]{Kiyeob Lee}

\author[1]{Ranyu Shi}

\author[2]{Pengwei Du}

\author[1]{Chanan Singh}

\author[1,3]{Le Xie}

\affil[*]{Corresponding author: \texttt{le.xie@tamu.edu}. A. Menati and X. Zheng are joint first author with equal contribution.}

\begin{abstract} 
Blockchain technologies are considered one of the most disruptive innovations of the last decade, enabling secure decentralized trust-building. However, in recent years, with the rapid increase in the energy consumption of blockchain-based computations for cryptocurrency mining, there have been growing concerns about their sustainable operation in electric grids. This paper investigates the tri-factor impact of such large loads on carbon footprint, grid reliability, and electricity market price in the Texas grid.
We release open-source high-resolution data to enable high-resolution modeling of influencing factors such as location and flexibility.
We reveal that the per-megawatt-hour carbon footprint of cryptocurrency mining loads across locations can vary by as much as 50\% of the crude system average estimate.
We show that the flexibility of mining loads can significantly mitigate power shortages and market disruptions that can result from the deployment of mining loads.
These findings suggest policymakers to facilitate the participation of large mining facilities in wholesale markets and require them to provide mandatory demand response.

\vspace{-2em}

\end{abstract}

\begin{document}
\flushbottom
\maketitle
\thispagestyle{empty}

\section{Introduction}
To grapple with the challenge of planetary-scale climate change, countries around the world are setting aggressive agendas towards carbon neutrality by mid 21st century, and electricity sector plays a pivotal role in achieving this goal.~\cite{WhiteHouse_crypto_report}
In the U.S., especially in Texas, there is a sharp increase of electrical loads that power blockchain-based proof-of-work computation of cryptocurrency mining. Specifically, between 2019 and 2022, U.S. hash rate share, the computational power required to mine new cryptocurrencies, increased notably from nearly 4 percent to 37 percent, and its power consumption increased from 300 MW to 5.7 GW.~\cite{Cambridge_index} A substantial percentage of these mining activities is happening in Texas, where there is currently more than 1.5 GW of mining capacity operating, and it is expected to attract nearly 2 GW of additional load per year.~\cite{LFLTF_collab_deck}
Such a rapid increase in the energy consumption of blockchain-based computations for cryptocurrency mining has raised concerns about their sustainable operation in electric grids.
Therefore, it is necessary to have a scientific, comprehensive understanding of the impact of mining loads.

\begin{figure}[hbtp]
\begin{minipage}{.45\textwidth}
\centering
\includegraphics[width=\textwidth]{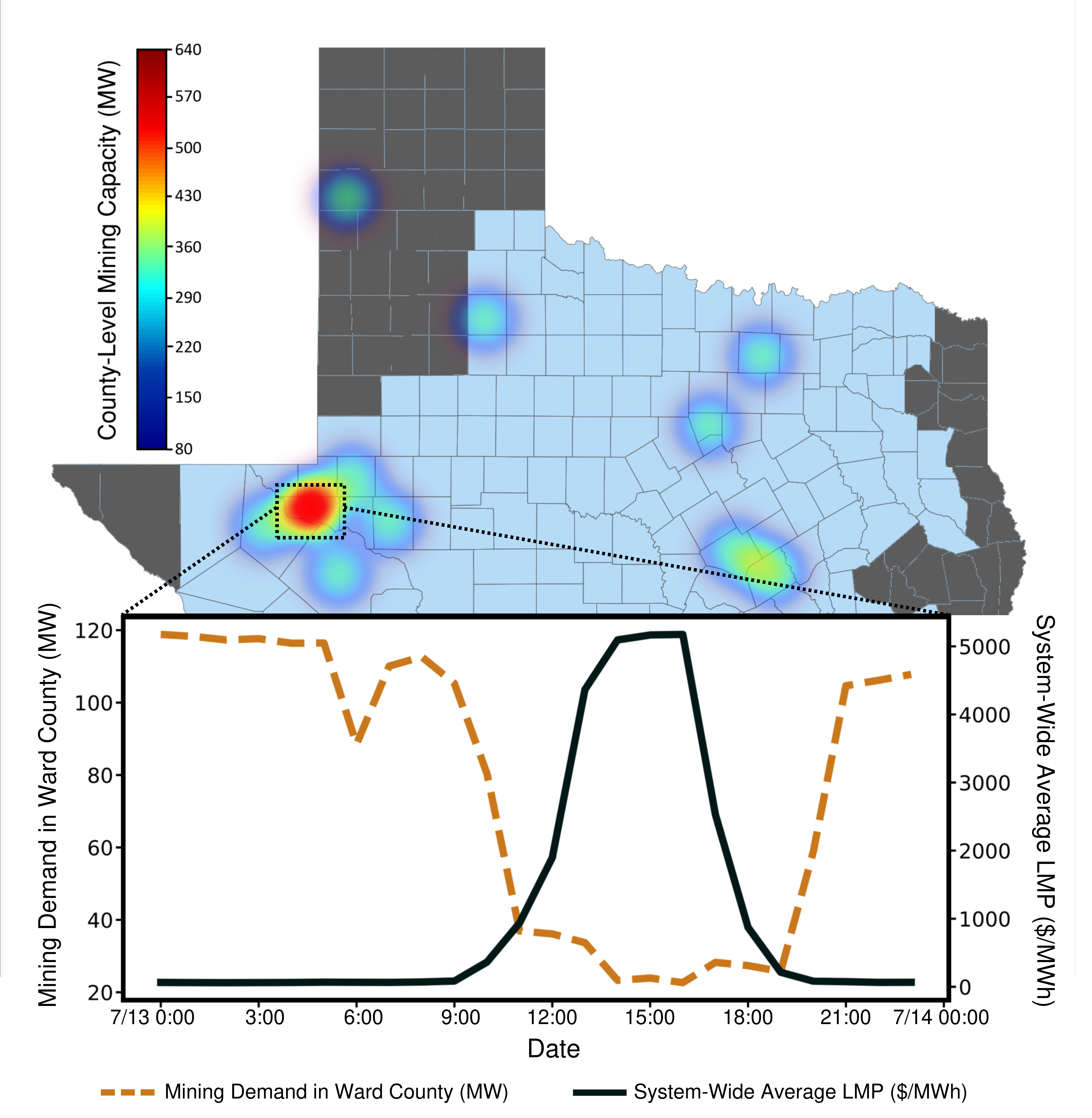}
\captionof{figure}{Negative correlation between real-world average LMP and mining load.}

\label{fig:mining_demand}
\end{minipage}\hfill
\begin{minipage}{.55\textwidth}
    \centering 
    \begin{threeparttable}
\captionof{table}{Correlation between total mining load, system-wide average LMP, and system-wide net load.}\label{tab:correlation}
\begin{tabular}{P{0.5cm} C{0.5cm} C{0.5cm}}
\toprule
\textbf{Correlation between} & \textbf{Whole period\tnote{1}} & \textbf{Summer peak\tnote{2}}  \\ \midrule
\textbf{Total mining load and system-wide average LMP} 
& -0.042 & -0.517 \\ \midrule
\textbf{Total mining load and system-wide net load} 
& 0.0667 & -0.757 \\ \midrule
\textbf{Non-mining load and system-wide average LMP} 
& 0.009 & 0.378 \\ \midrule
\textbf{Non-mining load and system-wide net load} 
& 0.922 & 0.971 \\ \bottomrule 
\end{tabular}
\begin{tablenotes}
    \item[1] The whole period refers to the period from January 1$^\text{st}$, 2021 to October 19$^\text{th}$, 2022.
    \item[2] The summer peak time refers to the period from July 7$^\text{th}$, 2022 to July 21$^\text{st}$, 2022.
  \end{tablenotes}
\end{threeparttable}
\end{minipage}
\end{figure}

While the area of digital assets and cryptocurrency are of high interest to society at large, its energy and environmental impact due to its high energy intensity remains largely unknown. 
 Most of the recent scholarly studies~\cite{krause2018quantification,Dittmar2019,rhodes2021impact, Masanet2019, stoll2019carbon,jones2022economic,niaz2022mining} and reports~\cite{Cambridge_index, WhiteHouse_crypto_report} draw their conclusions based on aggregated, low granularity data such as monthly or regional power consumption. 
This paper reveals for the first time higher-granularity data and provides insight that analyzes the impact of cryptocurrency mining on carbon footprint, system reliability, and market prices. To give a preliminary understanding of cryptocurrency mining loads, we analyze the correlation between cryptocurrency mining loads and external factors such as market price or electricity scarcity, using real data from the Texas grid (Fig.~\ref{fig:mining_demand} and Table~\ref{tab:correlation}). During the summer peak of 2022, the daily LMP fluctuated significantly, where the high demand during peak hours (2 - 5pm) created congestion in the system and led to large LMP spikes. Cryptocurrency mining loads showed strongly negative correlations with system-wide average local marginal price (LMP) and system-wide total load during these hours, while non-mining loads showed positive correlations. While it is difficult to deduce whether the root cause is mandatory demand response orders or spontaneous responses to high LMP, cryptocurrency mining loads are proved to be highly flexible, demonstrating the necessity of high-resolution modeling and analysis.

We summarize our main contributions as follows:

\begin{itemize}
    \item We develop an open-access tool that combines a large-scale grid model and high-resolution data to model cryptocurrency mining loads in grid operations under different scenarios, enabling a scientific, detail-oriented way to quantify the impact of cryptocurrency mining on the electricity sector.
    
    \item We study the location of cryptocurrency mining loads as a critical factor impacting carbon emissions. We show that the per-unit carbon footprint of mining loads exhibits locational disparity. Specifically, low-electricity-price locations can reduce per-unit carbon emission below 50\% of the system-wide average, while close-to-renewable locations do not necessarily lead to low carbon emission.
    
    \item We demonstrate that the flexibility of cryptocurrency mining loads plays a pivotal role in the reliability of electricity systems and the stability of electricity markets. We show that while the reliability of electric systems with higher renewable penetration is more susceptible to the integration of mining loads, full flexibility at all times can significantly avoid the reliability concerns created by the mining loads. We also show that a profit-driven price-responsive mining facility that only mines when the real-time local marginal price (LMP) is below \$40 per mega-watt-hour (MWh) can significantly mitigate extremely high LMPs across the system.
    
\end{itemize}

\section{Results}
\subsection{Per Unit Carbon Footprint of Cryptomining Has High Locational Disparity}

\begin{figure}[!tbp]
    \centering
    \includegraphics[width =1\textwidth]{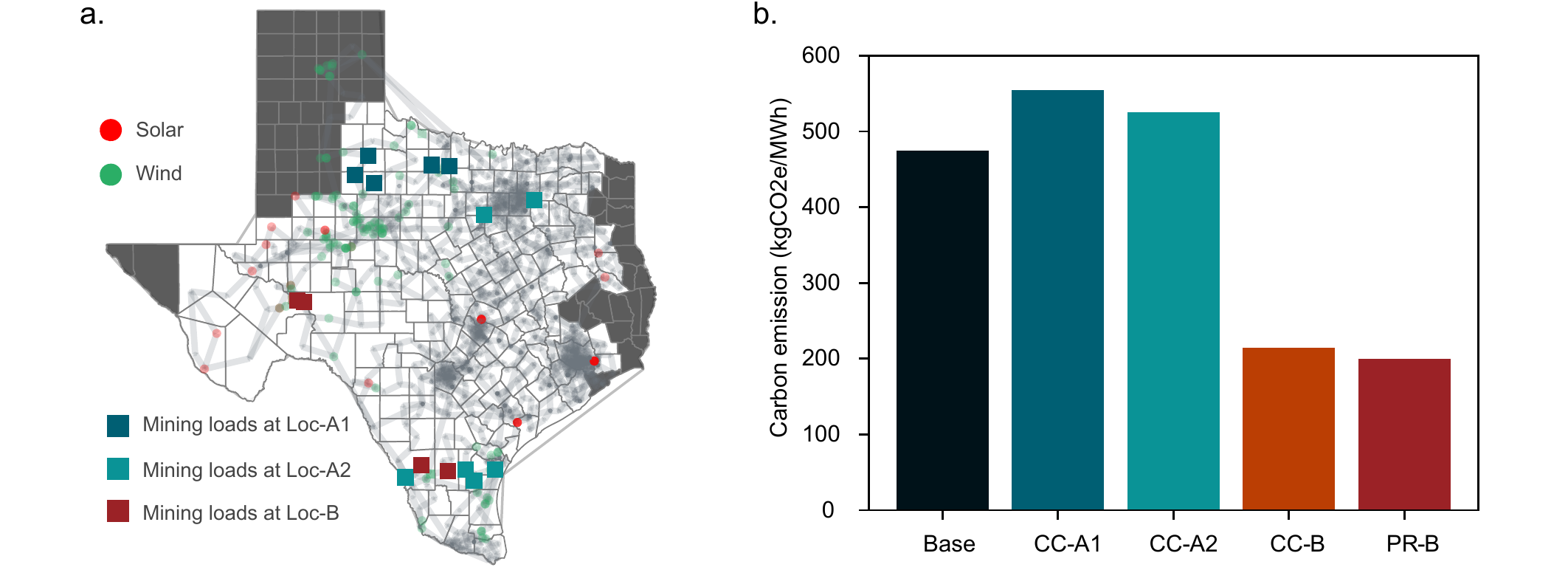}
    \caption{Carbon footprint per unit of cryptocurrency mining loads across the grid shows high inhomogeneity. a. Visualization of mining loads in synthetic Texas electricity grid, including close-to-renewable locations Loc-A1 and Loc-A2, and low-electricity-price locations Loc-B. b. Carbon footprint per MWh of total loads (Base), constant mining loads at Loc-A1 (CC-A1), Loc-A2 (CC-A2) and Loc-B (CC-B), and price-responsive mining loads at Loc-B (PR-B).}
    \label{fig:carbon}
\end{figure}
\begin{figure}[!tbh]
	\centering
	\includegraphics[width=1\textwidth]{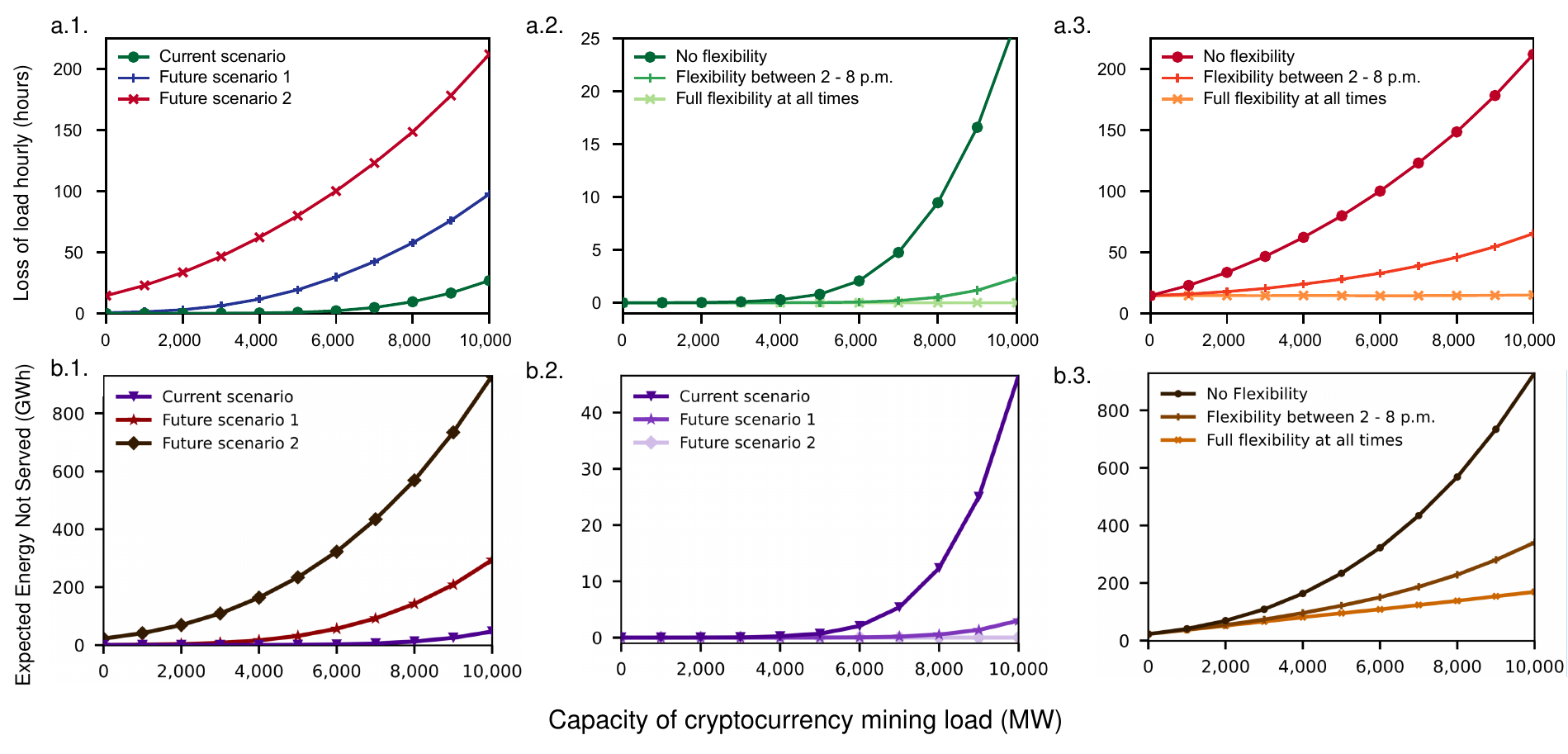}
	\caption{System reliability of different scenarios with different size and types of cryptocurrency mining loads as indicated by the loss of load hourly (a.1, a.2, a.3) and expected energy not served (b.1, b.2, b.3). 1. Comparison of the impact of non-flexible mining loads in the current scenario, future scenario 1 with 10\% more load and 50\% more renewables on the top of the current scenario, and future scenario 2 with 20\% more load and 100\% more renewables on top of the current scenario. 2. Comparison of different flexibility of mining loads in the current system scenario. 3. Comparison of the impact of mining loads with different flexibility in the future scenario 2.}
	\label{fig:reliability}
\end{figure}

The results suggest that the per unit carbon footprint of cryptocurrency mining has high heterogeneity across different locations in the synthetic Texas grid (Fig.~\ref{fig:carbon}). Here, the carbon footprint of cryptocurrency mining loads at a certain location is defined as the difference between the system-wide total carbon emission of the base system and that of the system with mining loads (see formal formulation in Section~\ref{sec:carbon_footprint_calculation}).
As shown in Fig.~\ref{fig:carbon}-b, adding new cryptocurrency mining loads in low LMP locations, such as Loc-B, could potentially lead to a per-unit carbon footprint below 50\% of the system average. On the other hand, integrating new loads in close-to-renewable locations, such as Loc-A1 and Loc-A2, can lead to per-MWh carbon footprints close to the system average. More details of location selection, flexibility mechanisms, and simulation process are available in Section~\ref{sec:experimental procedures}, and more results are available in Supplemental Figs. S1 and S2.
The results can be explained using a similar approach to the calculation of LMP.

Extensive research has shown that the LMP at a bus without local marginal generators is determined by an affine function of the marginal costs of generators located elsewhere. The coefficients in this function correspond to the change in active power of corresponding marginal generators in response to a hypothetical increase in load.~\cite{kirschen2018fundamentals} Furthermore, as per the definition of carbon footprint associated with increasing loads in Section~\ref{sec:carbon_footprint_calculation}, the carbon footprint is also an affine function of the changes in carbon footprint of marginal generators, with the same coefficients as those in the affine function of LMP. Additionally, the simulation model's generation bidding curves and life-cycle carbon emission factors demonstrate a strong correlation between a generator's marginal cost and its per-MWh life-cycle carbon emission.~\cite{Cambridge_index} Therefore, the LMP at a bus serves as a reliable indicator of local per-unit carbon emission.
On the other hand, as has been widely studied, the LMP at buses without local marginal generators can be higher, lower, or somewhere in between those at buses with marginal generators, which is not necessarily explicitly indicated by geographical locations. Observations through multi-day simulation also show that for most buses, including close-to-renewable buses, their LMPs are close to the system average, except for the selected low LMP locations. Therefore, close-to-renewable locations simply selected by geographic information-based criteria do not necessarily have low LMPs and hence do not necessarily indicate a lower per-unit carbon footprint.

The results indicate that conventional crude estimations of cryptocurrency mining's carbon footprint can be significantly off the mark. To accurately evaluate the carbon footprint of mining loads, it is necessary to incorporate location-in-the-grid. For example, the carbon emissions per MWh of individual miners and large-scale mining facilities may be quite different, because individual miners are typically concentrated in populated high-load areas where the LMP is typically relatively high, while large-scale mining facilities tend to be deployed in rural areas where LMPs are highly likely low. Spatial analysis of cryptocurrency miners' carbon footprints could also help policymakers determine proper carbon credit and tax policies to encourage miners in choosing locations with lower carbon emissions. Moreover, we hope this analysis will inspire researchers in the broader energy field to investigate how to scientifically calculate the carbon footprint of individual components in a networked system, where changes in individual components affect carbon emitters in complex ways.

\subsection{Flexibility Enhances System Reliability}
To capture the impact of cryptocurrency miners on the safe operation of the grid, we add different amounts of mining load into the system and plot the reliability index of Loss of Load Hourly (LOLH) and Expected Energy Not Served (EENS), which capture the expected number of hours per year that a system cannot serve load, and the expected energy that is not served, respectively. As shown in Fig.~\ref{fig:reliability} adding more mining loads increases both LOLH and EENS which leads to more frequent reliability incidents. In the coming years, with growing renewable sources on the grid, uncertainty and intermittency of the generation side will increase, making the system more susceptible to non-flexible loads. Demand flexibility is a potential solution to address the growing mismatch between the load and generation. Our results also suggest that in the future scenarios, both LOLH and EENS grow with a much faster rate compared to the current system (Fig.~\ref{fig:reliability}-a.1 and Fig.~\ref{fig:reliability}-b.1). The impact of mining loads is more detrimental in future scenario 2, with higher penetration of renewable energy and a larger increase in firm load (Fig.~\ref{fig:reliability}-a.1 and Fig.~\ref{fig:reliability}-b.1). More details of the future scenario creation and reliability assessment are available in the Experimental Procedure.

Cryptocurrency mining loads are profit-driven, and they tend to install substantial amounts of loads in a short period of time. Hence, it is difficult to perform conventional planning studies on time, and incentives new generators to cover the extra load on the system. The generation inadequacy can be solved by exploiting mining loads' flexibility and requiring them to operate during periods that will not compromise system reliability. 
As shown in (Fig.~\ref{fig:reliability}-a.2 and Fig.~\ref{fig:reliability}-b.2), without flexibility, adding mining loads threatens system reliability and significantly increases LOLH and EENS. On the other hand, complete or even partial flexibility, could potentially mitigate reliability incidents and completely offset the adverse impact of new mining loads.

It can be seen that for a system with 10 GW of added mining load, the LOLH and EENS are nearly 25 hours and 47 GWh per year, respectively. Which means that the detrimental impact of mining loads could be fully avoided without major economic loss (at most 25 hours of halted mining per year). In a future scenario (Fig.~\ref{fig:reliability}-a.3), this number increases to nearly 200 hours, which from a reliability perspective is a critical number, but from a cryptocurrency miner's point of view, it translates to not mining for 200 hours a year on average. Hence, cryptocurrency miner's demand flexibility is a win-win solution for them and the system operator.
Therefore, modeling cryptocurrency mining as “\emph{complete}” flexible demand is shown not to be detrimental to power grid reliability even with significant amounts at certain locations. Policymakers could consider these mining loads as virtual power plants capable of participating in the ancillary service market and account for their short- and long-term impacts on system reliability.

\subsection{Flexibility Mitigates Market Price Volatility}

Market clearing price in many wholesale electricity markets is typically determined by LMP. LMPs are characterized by statuses of marginal generating units and congestion of transmission lines. Delivering energy from a generator to a consumer at different location not only causes energy losses but also may overload the transmission lines, a phenomenon called congestion. LMPs tend to be volatile with the high price for the supply shortage and congestion in transmission line and almost zero price for the excessive supply, posing risks for the generation investment. The former is often observed during peak hours and the latter is often observed when abundant renewable energy exists in bulk power systems.

We investigate the roles of cryptocurrency mining loads on LMPs especially in terms of flexibility. We select three sets of locations on a synthetic Texas grid shown in Fig.~\ref{fig:carbon}-b (Loc-A1, Loc-A2, Loc-B) and investigate capacity, location and flexibility of mining loads in the 2000-bus system. Given both location and capacity of mining facilities determined, SCUC and SCED problems are solved to choose online/offline status of generating units and LMPs. Detailed simulation procedure is provided in Experimental Procedures.

\begin{figure}[!tbh]
    \centering
    \includegraphics[width=0.9\textwidth]{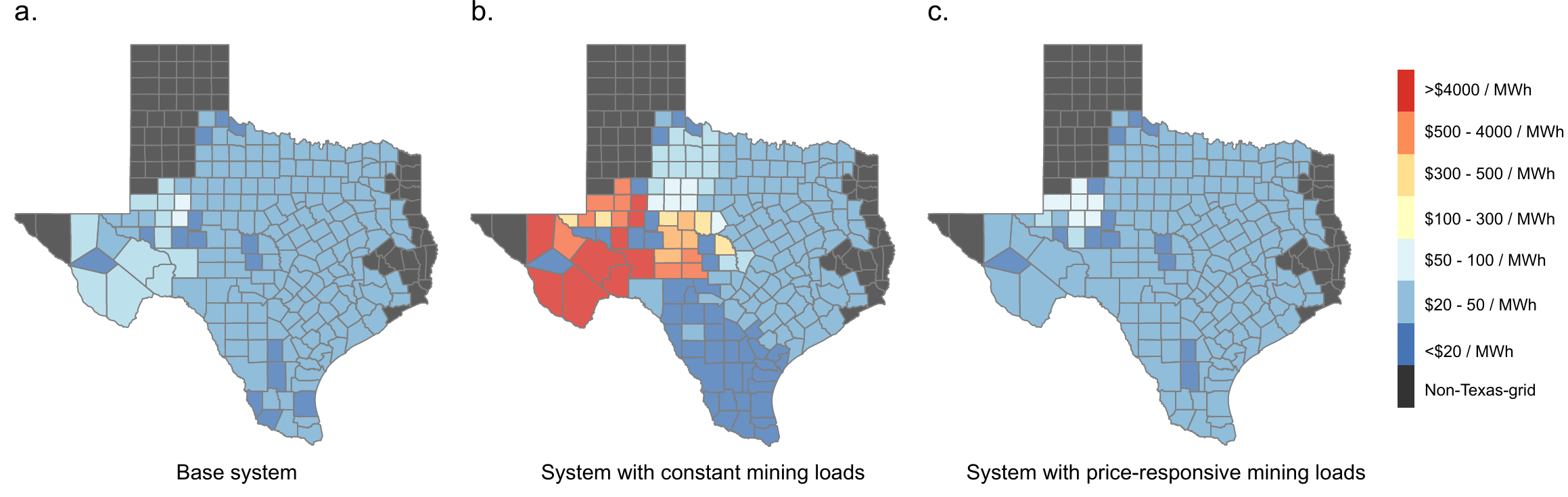}
    \caption{County-level LMP in the systems with or without different types of cryptocurrency mining loads. a. LMP in the base system without cryptocurrency mining loads. b. LMP in the system with constant cryptocurrency mining loads at Loc-B. c. LMP in the system with price-responsive cryptocurrency mining loads at Loc-B that only turn on mining when the real-time local marginal price is below \$40/MWh.}
    \label{fig:price}
\end{figure}

Fig.~\ref{fig:price} illustrates an instance of disruptions driven by mining loads in terms of price volatility. Fig.~\ref{fig:price}-a, b, c represent county-level LMPs for base case (without mining loads), constant mining loads, and price-responsive mining loads respectively. Observe that county-level LMPs (averaged LMP value for each county) for base case and price-responsive mining loads are almost identical that there only is negligible difference in LMPs whereas there are significant changes in Fig.~\ref{fig:price}-b. Fig.~\ref{fig:price} only shows an instance, yet we provide detailed statistics in Supplementary Fig. S~3. To remind the readers, any types of load growth will lead to the same disruptions while our focus is especially in terms of mining loads because they are growing in an unprecedented level. Therefore, based on our findings, it is essential to design new electricity market structures to maximally exploit the value of demand flexibility in mining loads. Price-responsive behaviour of these loads could potentially mitigate all their disruptive impact on the electricity market.

A summary of our results and observations is presented in Table~\ref{tab:result summary}. Future work will examine the impact of co-locating mining loads with renewable generation to further enable cryptocurrency miners' sustainable operation.

\begin{table}[ht]
\caption{\label{tab:result summary}Summary of cryptocurrency mining's impact on system reliability, carbon footprint, and electricity market.}
\begin{tabular}{c M{4cm} M{4cm} M{4cm}}
\toprule
& \textbf{System Reliability} & \textbf{Carbon Footprint} & \textbf{Electricity Market} 
\\ \midrule
\textbf{Flexibility} 
& Load flexibility could potentially avoid all reliability concerns without major economic loss
& The impact of demand flexibility on per MWh carbon emissions is not significant
& Price-responsive flexible operation of mining loads could mitigate market disruptions
 \\ \midrule
\\
\textbf{Location} 
& N/A  
& Low electricity price locations can control carbon emission below 50\% of system-wide average 
& Depending on the location, cryptominers impact on electricity market is highly non-uniform                   
\\ \bottomrule
\end{tabular}
\end{table}

\section{Conclusion}
This study uses high-resolution data to examine the effects of expanding cryptocurrency mining operations on the large-scale Texas power grid, focusing on carbon emissions, grid stability, and electricity prices. In contrast to previous research that used system-wide average carbon emission as a measurement, we found large variations in the carbon footprint of cryptocurrency mining across different locations, which can be as much as 50\% of the crude system average. Considering the fast integration of new mining facilities, offering proper financial incentives could potentially encourage the new facilities being built in locations with lower carbon footprint and high societal benefit.

Our findings indicate that the flexibility of cryptocurrency mining can greatly mitigate potential market disruption, while its impact on the electricity market can vary significantly depending on its location. This suggests that both location and flexibility are critical factors that policymakers and grid operators should consider when designing ancillary service programs that take advantage of the unique characteristics of the mining loads and enhance their grid-supporting capabilities. Furthermore, our analysis shows that treating cryptocurrency mining as fully flexible demand does not negatively impact grid reliability, even when present in large quantities at specific locations.

\section{Experimental Procedures}\label{sec:experimental procedures}

\subsection{{Data Processing}}
The data was obtained through the Electric Reliability Council of Texas (ERCOT) with permission to publish at county-level. The original data included the Supervisory Control and Data Acquisition (SCADA) measurements of twenty Large Flexible Loads (LFLs) within the ERCOT system at 5-minute intervals from January 1st, 2021 to December 31st, 2022. The LFLs span across nine counties and the average hourly total consumption in each county is calculated. One county's data showed zero-reading throughout the entire data set, therefore it was deleted from the final data set. During the data cleaning process, we noticed erroneous measurements that showed large spikes or negative values, these data points were filled with the previous value or set to zero, respectively. The county-level minimum mining capacity is 80 MW and the maximum mining capacity is 640 MW. However, it should be noted there is a large difference between the mining capacity and actual measured demand because many facilities are still in the process of expanding. 

\subsection{System Configuration}
In this paper, we quantify the impact of cryptocurrency mining loads with different flexibility at different locations in different system scenarios by simulating a large-scale synthetic Texas grid.
Therefore, this subsection will first describe the base system model, and then introduce the criteria for selecting cryptocurrency mining sites, designing flexibility mechanisms, and projecting future scenarios.

\subsubsection{Base Grid Model}

A calibrated large-scale synthetic Texas grid is used for optimal scheduling in this study,~\cite{xu2020us} which has been used in the studies of long-term planning,~\cite{xu2020us} extreme event analysis,~\cite{wu2021open,WU2022100106} and electricity market analysis.~\cite{lee2022targeted} This base grid model provides system topology, cost curves for fossil fuel generation units, renewable generation profiles, and base load profiles. More details of system calibration and data aggregation are described in the paper.~\cite{xu2020us}
For the forecasted load profiles, we aggregate zonal load profiles in 2020, and make appropriate adjustments (see Supplemental Note S2).

\subsubsection{Cryptocurrency Mining Deployment}
The step of cryptocurrency mining deployment aims to determine a set of buses with cryptocurrency mining loads $\mathcal{M}$ and capacity of cryptocurrency mining loads $\overline{P}^\text{m}$.
We consider four categories of cryptocurrency mining sites, including close-to-renewable, close-to-city, low-electricity-price, and real-mining sites.
For close-to-renewable sites, we manually select buses that are geographically close to solar or wind generation.
For close-to-city sites, we manually select buses that are geographically close to cities.
For low-electricity-price sites, we select buses with the lowest LMP under the condition of optimal scheduling of the base system that has $\mathcal{M}=\emptyset$ (see Model-based Simulation Process).
For real-mining sites, we select all buses in the counties where the LFLs recorded by the SCADA data are located.
For the former three categories, we define a uniform, constant capacity for all cryptocurrency mining loads. 
For real-mining sites, we define time-varying capacity of each mining load by scaling the LFL profile in the corresponding county to reflect the recorded dynamic power consumption of LFLs.
Visualization of cryptocurrency mining loads is shown in Supplemental Figure S1 and more details of settings are described in Supplemental Note S1.

\subsubsection{Flexibility Mechanism Design}
The flexibility mechanism determines actual power consumption profiles of cryptocurrency mining loads ${P}^\text{m}$, which will be added to load profiles in simulation.
We assume each cryptocurrency mining load has one binary state, namely, fully off or fully operational. Three types of flexibility mechanisms are considered in this study, including no flexibility (Eq.~\ref{eq:no_flexibility}), price-responsive flexibility (Eq.~\ref{eq:price_response_flexibility}), and command-following flexibility (Eq.~\ref{eq:command_following_flexibility}). Specifically, the power consumption of cryptocurrency mining loads with different flexibility is defined by
\begin{subequations}
\begin{align}
    &{P}^\text{m}_{ihd} = \begin{cases}
        \overline{P}^\text{m}_{i} & \text{if } i\in \mathcal{M}\\
        0 & \text{otherwise}\\
    \end{cases}\label{eq:no_flexibility}\\
    &{P}^\text{m}_{ihd} = \begin{cases}
        \overline{P}^\text{m}_{i} & \text{if } i\in \mathcal{M} \text{ and } {\mu}_{ihd}\leq \overline{\mu}\\
        0 & \text{otherwise}\\
    \end{cases}\label{eq:price_response_flexibility}\\
    &{P}^\text{m}_{ihd} = \begin{cases}
        \overline{P}^\text{m}_{i} & \text{if } i\in \mathcal{M} \text{ and } \text{ not } \left(\mathbbm{1}^\text{ctx}_{ihd} \text{ and } \mathbbm{1}^\text{cmd}_{ihd}\right)\\
        0 & \text{otherwise}\\
    \end{cases}\label{eq:command_following_flexibility}
\end{align}
\end{subequations}
where
${P}^\text{m}_{ihd}$ is the active power consumption of the cryptocurrency mining load at bus $i$ at hour $h$ on day $d$,
$\overline{P}^\text{m}_{i}$ is the installed capacity of the cryptocurrency mining load at bus $i$,
$\mathcal{M}$ is a set of buses with cryptocurrency mining loads,
${\mu}_{ihd}$ is the LMP at bus $i$ at hour $h$ on day $d$,
$\overline{\mu}$ is a pre-defined threshold that is set as \$40 /MWh in this study,
$\mathbbm{1}^\text{cmd}_{ihd}$ indicates whether system operators order a shutdown at the given moment,
and $\mathbbm{1}^\text{ctx}_{ihd}$ indicates whether the cryptocurrency mining load provides demand response services at the given moment.

These flexibility mechanisms are designed considering real-world demand response programs currently adopted by cryptocurrency mining facilities. As it can be seen, in the price-responsive flexibility (Eq.~\ref{eq:price_response_flexibility}) cryptominers stop mining if the electricity prices are greater than a certain threshold ($\overline{\mu}$). This threshold either depends on their operational profit and their break-even electricity price, or it is determined by a demand response contract with their electric utility company. Another practical flexibility mechanism is command-following (Eq.~\ref{eq:command_following_flexibility}), in which the mining loads receive a command from the grid operator to shut down their load, and depending on their obligation, they might decide to keep mining or reduce their load as requested by the operator.

\subsubsection{Future Scenario Creation}
To compare the impact of cryptocurrency mining on system reliability, we study the current synthetic grid and a future grid representing the generation mix change in Texas. Currently, the peak load in Texas is nearly 77 GW, but this load is predicted to reach 87 GW to 93 GW by 2030, depending on the potential of load reduction programs.~\cite{CDR} This increase in firm load is nearly 10\% to 20\% of the current load. At the same time, the generation mix is undergoing substantial change and it is moving toward a highly renewable grid. While the traditional gas, coal, and nuclear generator capacities will not see a major change, the renewable generation capacity is expected to grow substantially. By 2030, the amount of renewable generation is planned to increase from 17,000 MW to nearly 41,500 MW.~\cite{CDR} Hence, we project 50\% to 100\% of potential increase in the renewable generation capacity in Texas. Therefore, two future scenarios are designed in our simulations. In scenario 1, the synthetic Texas grid's base load is increased by 10\% and the renewable generation capacity is increased by 50\%, and in scenario 2, the base load is increased by 20\%  and the renewable generation capacity is increased by 100\%.

\subsection{{Model-based Simulation Process}}
The process of optimal scheduling consists of alternating security-constrained unit commitment (SCUC) and security-constrained economic dispatch (SCED), mimicking day-ahead and real-time electricity market clearing processes. SCUC is used for day-ahead market clearing and aims to find optimal commitments and dispatch decisions based on day-ahead load and renewable forecasts. SCED is used for real-time market clearing and aims to find more accurate dispatch decision based on solved optimal commitments and accurate short-term load and renewable forecasts.
Please refer to the previous study~\cite{lee2022targeted} for detailed mathematical formulation of SCUC and SCED.

It is worth noting that we make a few assumptions and simplifications for simulation.
First, we assume high accuracy of load and renewable generation forecasts. Therefore, SCUC and SCED share the same deterministic load and renewable generation profiles in this study.
Second, we only consider cryptocurrency mining loads with no flexibility or price-response flexibility for simulation.
Third, we assume that the generation states in an optimal UC solution are not affected by the mining load behaviour, so for simplicity assume no flexibility for all mining loads for SCUC.
Therefore, SCUC and SCED have one key difference in this study. Only the generation output $\{P^{\text{g}}_{ihd}\}$ and LMP $\{\mu_{ihd}\}$ solved by SCED are used for the analysis of carbon footprint and market prices, while those solved by SCUC are only a reference for cryptocurrency mining loads with price-response flexibility.

We design and implement Algorithm~\ref{algo:simulation} by MATPOWER optimal scheduling tool (MOST),~\cite{zimmerman2016matpower} where $\mathcal{G}$ represents a grid model, $\mathcal{M}$ is a set of buses with cryptocurrency mining loads, $\overline{P}^\text{m}$ is installed capacity of all cryptocurrency mining loads, ${P}^\text{l}$ is base load profiles, and $D_1$ and $D_T$ are the start and end dates of the period of interest. In summary, through the process of optimal scheduling, we can finally obtain hourly LMP $\{\mu_{ihd}\}$, generation output $\{P^\text{g}_{ihd}\}$ and cryptocurrency mining power consumption $\{{P}^\text{m}_{ihd}\}$.
It is worth noting that a minor fraction of days in some scenarios is excluded due to infeasible SCUC solutions (see Supplemental Note S2).

\begin{algorithm}[htpb!]
\caption{Process of optimal scheduling by alternating SCUC and SCED}
\label{algo:simulation}
    \begin{algorithmic}
        \State \textbf{Input:} $\mathcal{G}$, $\mathcal{M}$, $\overline{P}^\text{m}_{i}$, ${P}^\text{l}_{ihd}$, $D_1$, $D_T$
        \State \textbf{Output:} $\{P^{\text{g}}_{ihd}\}$, $\{\mu_{ihd}\}$, $\{{P}^\text{m}_{ihd}\}$
        \For{$d=D_1,\cdots,D_T$}
            \State \textbf{Initialize} $U^{\text{g}0}_{i}\leftarrow U^\text{g}_{i\left(24\right)\left(d-1\right)},$ $i=1,\cdots,N$, \Comment{Start day-ahead market clearing by SCUC}
            \State \textbf{Initialize} ${P}^\text{m}_{ihd}$ based on Eq.~\ref{eq:no_flexibility} given $\mathcal{M}$ and $\overline{P}^\text{m}_{i},$  $i=1,\cdots,N$, $h=1,\cdots,24$,
            \State \textbf{Calculate} ${P}^\text{ttl}_{ihd}\leftarrow {P}^\text{m}_{ihd}+{P}^\text{l}_{ihd},$  $i=1,\cdots,N$, $h=1,\cdots,24$, 
            \State $\{U^{\text{g}}_{ihd}\}_{h=1}^{24}$, $\{P^{\text{g}}_{ihd}\}_{h=1}^{24}$, $\{\mu_{ihd}\}_{h=1}^{24}$ $\leftarrow$ $\text{SCUC}\left(\mathcal{G}, U^{\text{g}0}_{i}, \{{P}^\text{ttl}_{ihd}\}_{h=1}^{24}\right)$
            \If{price-response flexibility} \Comment{Start real-time market clearing by SCED}
                \State \textbf{Update} ${P}^\text{m}_{ihd}$ based on Eq.~\ref{eq:price_response_flexibility} given $\mathcal{M}$, $\overline{P}^\text{m}_{i}$, and $\{\mu_{ihd}\}_{h=1}^{24},$ $i=1,\cdots,N$, $h=1,\cdots,24$,
                \State \textbf{Update} ${P}^\text{ttl}_{ihd}\leftarrow {P}^\text{m}_{ihd}+{P}^\text{l}_{ihd},$ $i=1,\cdots,N$, $h=1,\cdots,24$,
            \EndIf
            \State $\{P^{\text{g}}_{ihd}\}_{h=1}^{24}$, $\{\mu_{ihd}\}_{h=1}^{24}$ $\leftarrow$ $\text{SCED}\left(\mathcal{G}, \{U^{\text{g}}_{ihd}\}_{h=1}^{24}, \{{P}^\text{ttl}_{ihd}\}_{h=1}^{24}\right)$
        \EndFor
    \end{algorithmic}
\end{algorithm}

\subsection{Carbon Footprint Calculation}\label{sec:carbon_footprint_calculation}
We define the carbon footprint of cryptocurrency mining loads $C^\text{m}$ as the difference between the system-wide total carbon emission of the base system $C$ and that of the system with mining loads $\hat{C}$ by
\begin{subequations}
\begin{align}
    &C = \sum_{d=D_1}^{D_T}\sum_{h=1}^{24}\sum_{i=1}^{N} P^\text{g}_{ihd}\cdot \phi_i, \\
    &\hat{C} = \sum_{d=D_1}^{D_T}\sum_{h=1}^{24}\sum_{i=1}^{N} \hat{P}^\text{g}_{ihd}\cdot \phi_i, \\
    &C^\text{m} = C - \hat{C},
\end{align}
\end{subequations}
where $P^\text{g}_{ihd}$ is the active power output of generator $i$ at hour $h$ on day $d$ in the base system, $\hat{P}^\text{g}_{ihd}$ is the active power output of generator $i$ at hour $h$ on day $d$ in the system with mining loads,
$\phi_i$ is the life-cycle greenhouse emission factors of generator $i$ determined by the generation type,~\cite{Cambridge_index} $N$ is the total number of buses, and $D_1$ and $D_T$ are the start and end dates of the period of interest.

Further, we define the per unit energy consumption carbon footprint of system loads $\Bar{C}$ and cryptocurrency mining loads $\Bar{C}^\text{m}$ by
\begin{subequations}
\begin{align}
    \Bar{C} &= C/\left(\sum_{d=D_1}^{D_T}\sum_{h=1}^{H}\sum_{i=1}^{N} P^\text{l}_{ihd}\right), \\
    \Bar{C}^\text{m} &= C^\text{m}/\left(\sum_{d=D_1}^{D_T}\sum_{h=1}^{H}\sum_{i=1}^{N} \hat{P}^\text{m}_{ihd}\right),
\end{align}
\end{subequations}
where $P^\text{l}_{ihd}$ is the active power consumption of the load at bus $i$ at hour $h$ on day $d$ in the base system, and $\hat{P}^\text{m}_{ihd}$ is the active power consumption of the cryptocurrency mining load at bus $i$ at hour $h$ on day $d$ in the system with mining loads.

Comparing to methods that estimate carbon footprint based on the average generation mix, this approach can incorporate the nonlinear spatial-temporal relation between loads and generators in the process of optimal scheduling.

\subsection{{Reliability Assessment}}
For reliability assessment, we use the Monte-Carlo method to simulate the randomness in the failure and repairs of the generators. In the Monte-Carlo method, each generator has a Mean Time to Failure (MTTF), which is the expected number of hours the generator will operate before it fails, and a Mean Time to Repair (MTTR), which is the expected number of hours it takes to bring the generator back into operation. These numbers are obtained from the standard IEEE Reliability Test System.\cite{rts} The failure (repair) time of each generator is randomly generated from an exponential distribution with MTTF (MTTR) as its mean value. We do not consider a failure rate for renewable generators, but their output is fluctuating and highly uncertain. Using the generation and load profiles obtained from the synthetic grid, we repeat the simulation 10,000 times until the reliability indices start to converge.

We calculate the Loss of Load Hourly (LOLH) as the reliability/adequacy index, which captures the expected number of hours per year that the system cannot serve load. The LOLH is calculated for the current synthetic grid and two future scenarios, where in scenario 1 (scenario 2), today's synthetic grid is updated by increasing the load by 10\% (20\%) and increasing the renewable generation capacity by 50\% (100\%).

\subsection{Market Price Analysis}
For ease of visualization, given hourly nodal LMP $\{\mu_{ihd}\}$, we define the average LMP in county $k$ by
\begin{equation}
    \Bar{\mu}^{\text{cty}}_{khd} = 1/|z_k| \cdot \sum_{j\in z_k}\mu_{jhd},
\end{equation}
where $z_k$ is the set of buses in county $k$.

\section{Data and Code Availability}
The data, model and codes for all the analyses in this paper are publicly available at the Github repository~\url{https://github.com/tamu-engineering-research/Crypto_mining_impacts}.

\Urlmuskip=0mu plus 1mu\relax
\bibliography{ref}

\newpage

\section{Supplemental Figures}
\subsection{Figure S1 Visualization of renewables and cryptocurrency mining loads}
Loc-A1 and Loc-A2 are selected based on the criteria of close-to-renewable sites. Loc-B is selected based on the criteria of low-LMP sites. Loc-C is selected based on the criteria of close-to-city sites. Loc-D includes all buses in the counties where real mining loads are located.
\begin{figure}[htbp]
    \centering
    \includegraphics[width =0.7\textwidth]{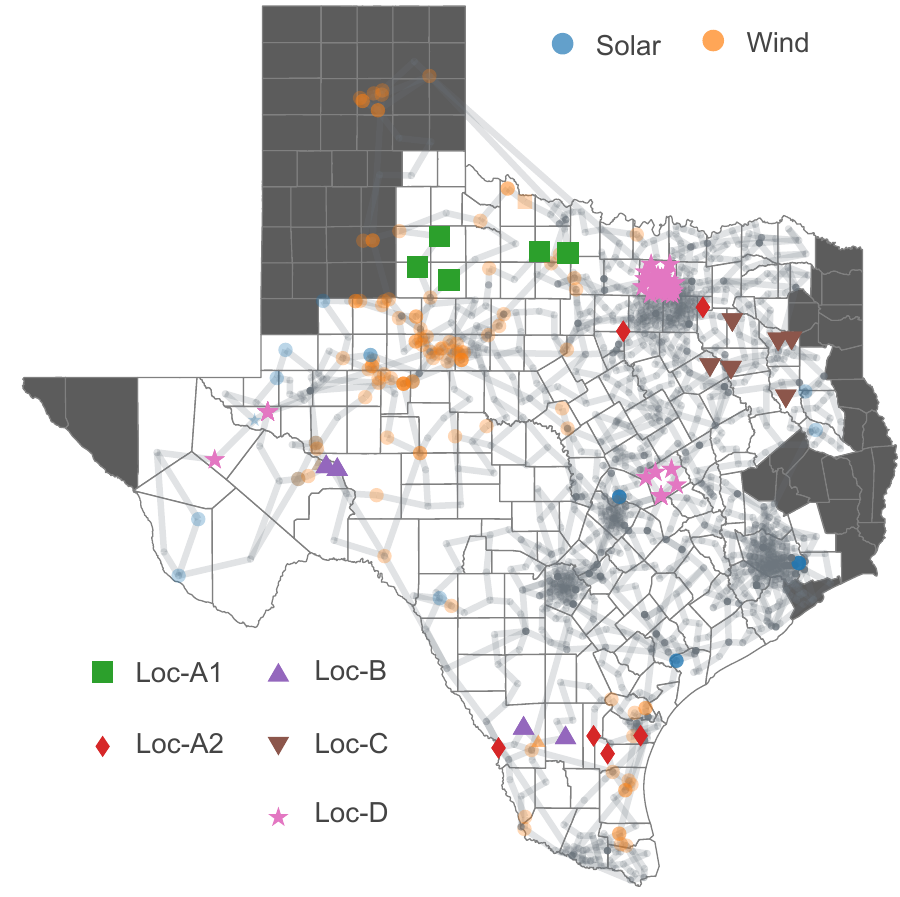}
    \caption{Visualization of renewables and cryptocurrency mining loads}
    \label{fig:mining_position}
\end{figure}

\clearpage

\subsection{Figure S2 Carbon footprint per unit of electricity consumption of cryptocurrency mining loads at different locations}
Loc-A1 and Loc-A2 are selected based on the criteria of close-to-renewable sites. Loc-B is selected based on the criteria of low-LMP sites. Loc-C is selected based on the criteria of close-to-city sites. Loc-D includes all buses in the counties where real mining loads are located. CC means constant mining loads, while PR means price responsive mining loads that only mine when real-time LMP is below \$40 /MWh.
\begin{figure}[htbp]
    \centering
    \includegraphics[width =0.8\textwidth]{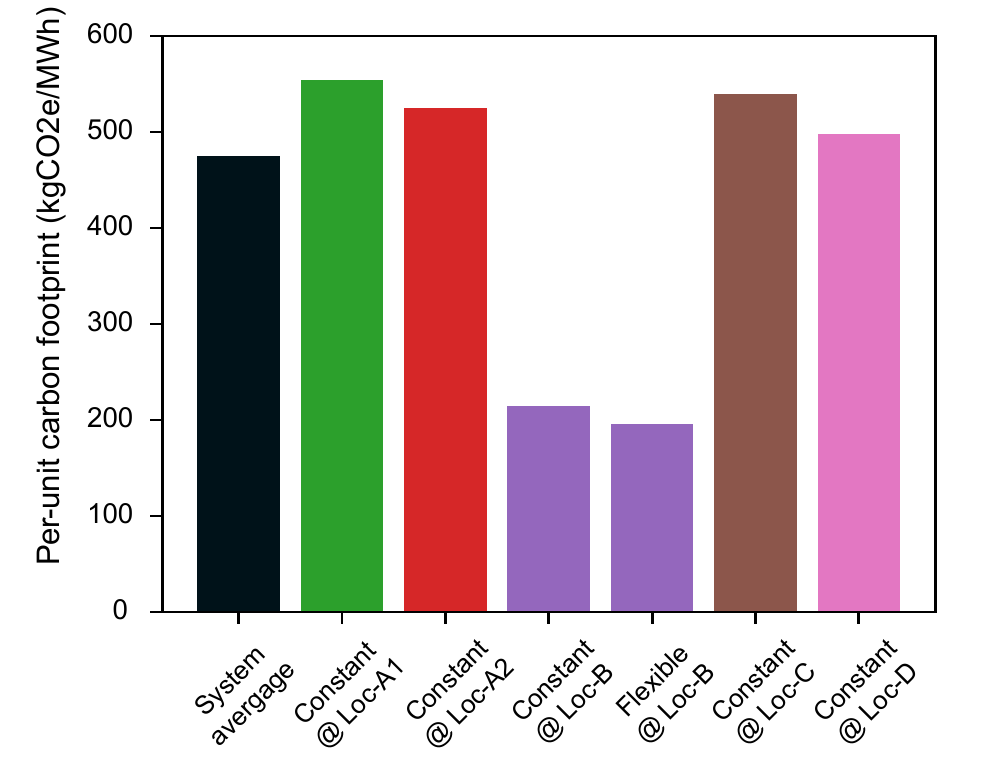}
    \caption{Carbon footprint per unit of electricity consumption of cryptocurrency mining loads at different locations}
    \label{fig:carbon-2}
\end{figure}
\clearpage

\subsection*{Figure S3 Statistics of LMPs in Summer 2020}
\begin{figure}[htbp]
    \centering
    \includegraphics[width =0.6\textwidth]{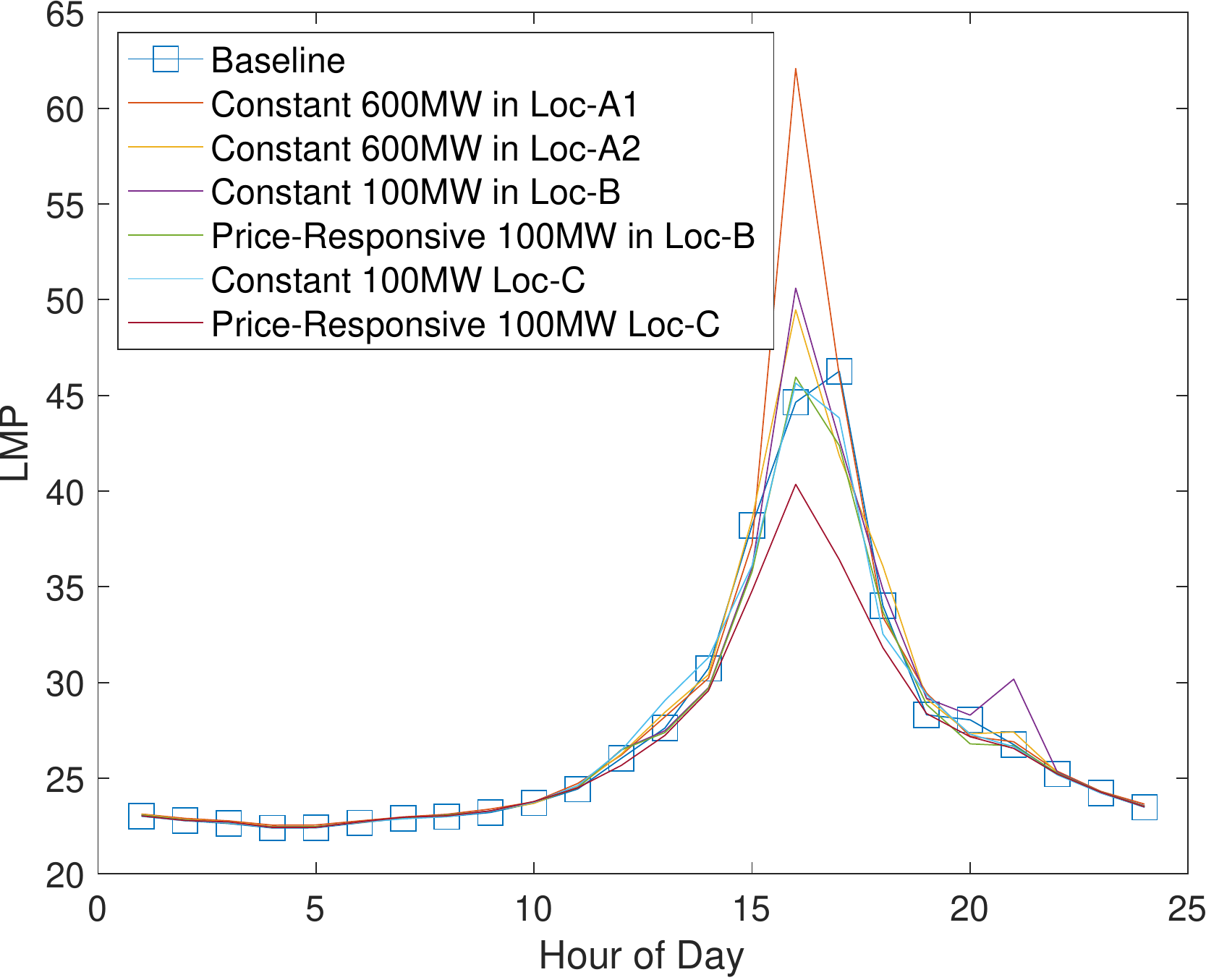}
    \caption{Hourly LMPs}
    \label{fig:statistics-LMPs.pdf}
\end{figure}

Statistics of hourly LMPs are shown in Fig.~S \ref{fig:statistics-LMPs.pdf}. Our experiments are 61-days (June 28$^\text{th}$ 2020 to August 27 $^\text{th}$ 2020) and hourly LMPs are averaged values for all hours over 61-days. Fig.~S \ref{fig:statistics-LMPs.pdf} includes a number of scenarios where we assume a total capacity of 600 MW (or 100 MW) is installed in selected locations such as Loc-A1, Loc-A2, Loc-B, Loc-C. Also, there are two operational modes, namely constant consumption and price-responsive modes. Fig.~S \ref{fig:statistics-LMPs.pdf} shows that LMPs could become significantly higher when constant mining loads are added to the synthetic Texas grid, however, LMPs may not become significantly higher when the loads are price-responsive.

\clearpage

\subsection*{Figure S4 Seven days county-level real-world mining load in Ward County}
\begin{figure}[htbp]
    \centering
    \includegraphics[width =0.6\textwidth]{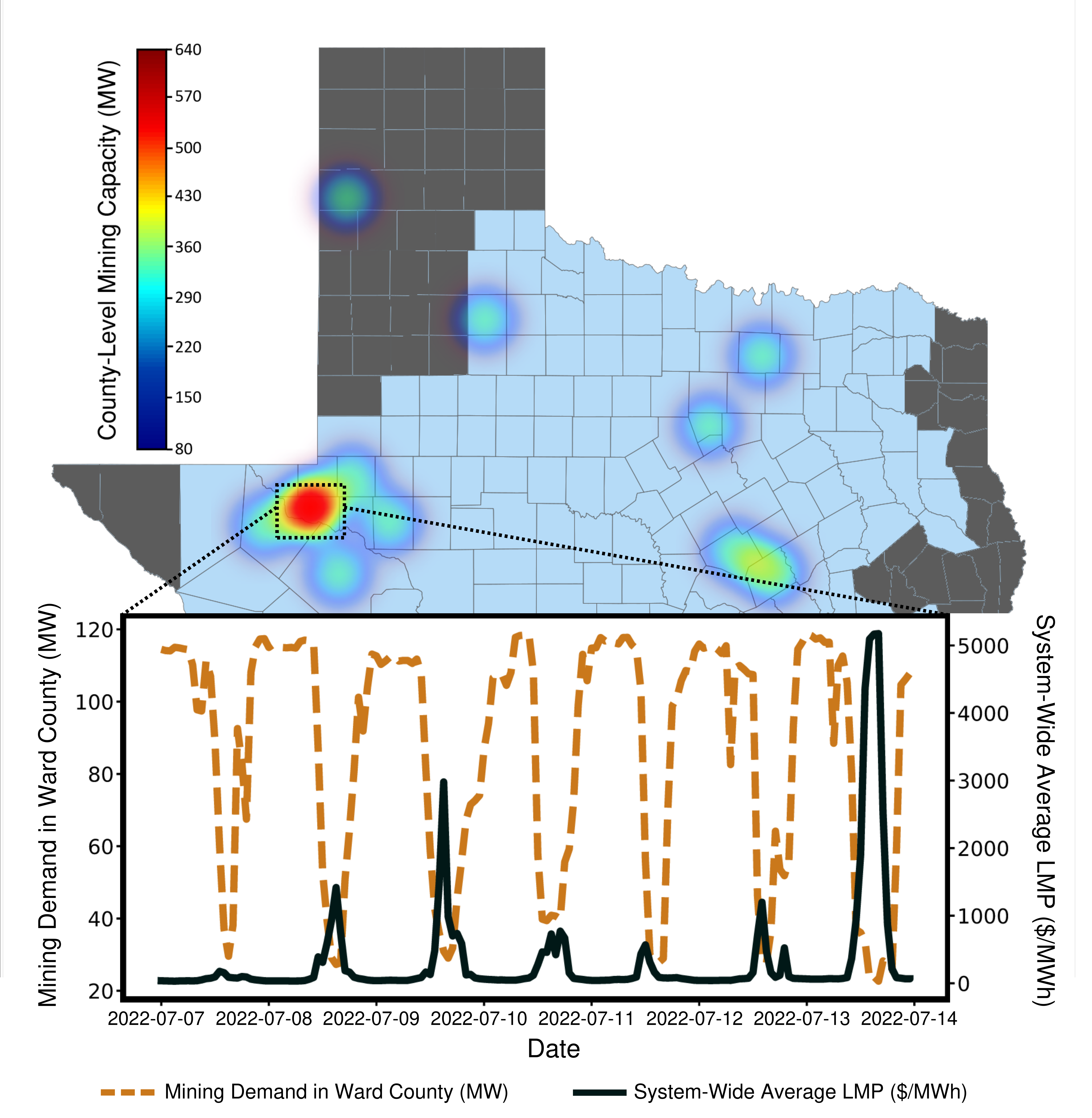}
    \caption{Mining Load in Ward County from 7/7/2022 - 7/13/2022}
    \label{fig:7day}
\end{figure}

Peak LMP day and corresponding mining demand in Ward county is shown in Fig. S \ref{fig:mining_demand}. We also include seven days(July 7$^\text{th}$ 2022 - July 13$^\text{th}$ 2022) of demand data in Ward county in Fig. \ref{fig:7day} to show repeated flexibility from cryptocurrency mining demand through out seven days.

\section*{Supplemental Notes}
\subsection*{Note S1 Details of mining locations}
The set of cryptocurrency mining loads $\mathcal{M}$ of different selection criteria is defined as follows:
\begin{itemize}
    \item Close-to-renewable locations Loc-A: \{122, 140, 157, 159, 178, 202\}
    
    \item Close-to-city locations Loc-C: \{569, 1895, 1896, 1963, 1948, 1994\}
    \item Real locations Loc-D: \{13, 47, 48, 49, 50, 51, 52, 53, 572, 577, 582, 643, 660, 664, 678, 725, 729, 730, 731, 732, 740, 753, 760, 780, 783, 785, 789, 818, 819, 842, 872, 880, 906, 907, 908, 935, 987, 988, 989, 1029, 1030, 836, 1157, 1158, 1159, 1160, 1161, 1177, 1375, 1388, 31\}
    \item Low-LMP locations Loc-B: \{8, 9, 58, 69, 70, 499, 545, 546, 547\}
    \item Close-to-renewable locations Loc-E: \{402, 403, 493, 497, 743, 903\}
\end{itemize}

\subsection*{Note S2 Infeasible days to solve optimal scheduling}
In this paper, we focus on analyzing the impacts of cryptocurrency mining loads in the period from Day 180 (June 28$^\text{th}$) to Day 240 (August 27$^\text{th}$) in summer 2020. Typically, the summer period has high demand in ERCOT markets strongly correlated with hot weather in the summer. A demand profile on the synthetic Texas grid is designed similarly, so it has high demand in the summer period as well, which is the most interesting time period to study. Additionally, we uniformly reduce the system-wide load during the period from Day 221 (June 28$^\text{th}$) to Day 224 (June 28$^\text{th}$) by 4\% in order to obtain feasible optimal scheduling solutions.

\end{document}